\documentstyle[prl,aps,preprint,tighten,floats,epsfig]{revtex}

\renewcommand{\baselinestretch}{1.2}

\def\btabl{\begin{table}}   \def\etabl{\end{table}}
\def\bea{\begin{eqnarray}}   \def\eea{\end{eqnarray}}
\def\bnn{\begin{eqnarray*}}   \def\enn{\end{eqnarray*}}
\def\beq{\begin{equation}}   \def\eeq{\end{equation}}  
\def\btabu{\begin{tabular}}   \def\etabu{\end{tabular}}
\def\bec{\begin{displaymath}} \def\eec{\end{displaymath}}
\def\nn{\nonumber}
\def\eqref#1{(\ref{#1})}
\newcommand{\sumint}{\sum\!\!\!\!\!\!\!\int}

\begin{document}
\draft
%\date{\today}
\preprint{\vbox{\baselineskip=13pt
\rightline{CERN-TH/98-309}
%\vskip0.2truecm
\rightline{FAMNSE-98/37 }
%\vskip0.2truecm
\rightline{LPTHE Orsay-98/60}
\vskip0.2truecm
\rightline{hep-ph/9810449}}}
%%\twocolumn[\hsize\textwidth\columnwidth\hsize\csname
%%@twocolumnfalse\endcsname
\title{Addendum to Finite-size effects on multibody neutrino exchange}
\author{A. Abada} 
\address{{\small Theory Division, CERN, CH-1211 Geneva 23, Switzerland\\asmaa.abada@cern.ch}}
\author{O. P\`ene and J. Rodr\'\i
guez-Quintero} 
\vskip -0.3cm
\vspace{-0.5cm}
\address{{\small 
Laboratoire de Physique Th\'eorique et Hautes
Energies\footnote{Laboratoire associ\'e Centre National de la Recherche Scientifique -
URA D00063.}, \\
Universit\'e de Paris XI, B\^at 211, 91405 Orsay Cedex,
France\\pene@qcd.th.u-psud.fr, jquinter@qcd.th.u-psud.fr}}

\vskip -1cm
\maketitle \begin{abstract}  
\vskip -0.5cm
The interaction energy of the neutrons due to
 massless neutrino exchange in a neutron
star has recently been proved, using an effective theory, 
to be extremely small and infrared-safe. 
Our comment here is of conceptual order: 
two approaches to compute the total interaction
energy density have  recently  been proposed. 
Here, we study the connection between these two approaches.
 From CP invariance, we argue that the resulting 
interaction energy  has  to be even in
the parameter $b=-G_F n_n /\sqrt{2}$, which expresses the static
neutrino  potential created by a neutron medium of density $n_n$. 

\medskip

\end{abstract}
\vskip 1.0cm
%%\vskip 1.0cm]

\leftline{CERN-TH/98-309}
\leftline{October 1998}

\renewcommand{\baselinestretch}{1.2}
\newpage

The long-range neutrino-mediated interaction between neutrons in a 
dense core,
 such as a neutron star, has  recently been studied 
 \cite{rescue,kachelriess,note,kiers,border,mimura} 
and shown to be extremely small. 
Following the method of Schwinger \cite{schwinger}, the interaction energy 
$W$ can be computed as
\bea
 W=\langle\hat{0}|{ H}|{\hat{0}}\rangle - 
 \langle{0}|{ H}_0|{0}\rangle\ ,\label{W}
 \eea  
 where $|{\hat{0}}\rangle, |{{0}}\rangle$
 are the matter and
 matter-free vacua,  respectively, and $H$ ($H_0$) is the Hamiltonian 
 in the presence (in the absence) of 
 the star medium. For convenience we will call, in this note, ``matter vacuum'' 
 the neutrinoless ground state in the presence of a star. 
 This first step is crucial: {\it the interaction energy is equal to the
 shift of the zero-point energy of a neutrino due to the presence of the star.}
We have demonstrated \cite{note,border} that there is a non-vanishing
 zero-point energy density difference ($W$) between the inside and the 
 outside
  of the star, which is due to the refraction index at the stellar
   boundary and the resulting
non-penetrating waves. In Ref. \cite{border}, this effect  was shown 
analytically and numerically to be the
dominant one and lead to an infrared-safe total energy density.
 This result is in contradiction with the previous claim  in
  \cite{fischbach}
  that there must be a lower bound 
on the neutrino mass to ensure the existence of stars. The latter
 ``catastrophic result" is a
consequence of summing up large infrared terms outside the radius 
of convergence of the
perturbative series. 
The use of an effective Lagrangian (\ref{lageff}), which can
 be exactly
solved, allows this non-convergence to be circumvented 
 and to provide the result to a good accuracy.
The effective Lagrangian writes:
 \bea
 {\cal {L}}_{\mathrm{eff}}=i{\overline{\nu}_L} \partial\!\!\!/ \ \nu_L ( r) 
 -b\ { \overline {\nu}_L}\gamma_0\nu_L \ \theta(R -r )\ ,\label{lageff}
 \eea
where $R$ is the radius of the neutron star and $b=-G_F n_n /\sqrt{2}$ 
 summarizes the static potential felt by the neutrinos, 
 which is generated by the neutrons through $Z_0$-exchange.
 The neutrons are treated as static, and
  represented by a uniform axial-charge density
  (being electrically neutral the neutrons interact
  with the $Z_0$ only the via axial current) while $n_n$
   is the star neutron density. Notice  that 
 the potential is attractive, 
 with depth $b$,  for neutrinos, which then condense,
  and repulsive for antineutrinos. It is important to stress that
  this effective Lagrangian, Eq. (\ref{lageff}), is non-CP-invariant since it
  describes the neutrino in a non-CP-symmetric background: 
  the neutron star. In fact, it is easy to see that
 $ CP \ {\cal {L}}(b) \ (CP)^{-1} \ = \ {\cal {L}}(-b)$.

 As a consequence of the 
 effective theory used, 
  expression (\ref{W}) is formally an ultraviolet-divergent quantity,
 which needs to be regularized. 
 This can be written as

\beq
W=\sum_{i<0} E_i-E^0_i, \label{vide1}
\eeq

\noindent where $i$ runs over the negative neutrino energy levels.
 Another method \cite{kiers,border}  consists 
 in the following symmetrization: 
 
\beq
W_{\mathrm{sym}}=-1/2\left(\sum_{i>0}( E_i-E^0_i)-\sum_{i<0}( E_i-E^0_i)\right). 
\label{vide2}
\eeq

\noindent As a matter of fact, Eqs. (\ref{vide1}) and (\ref{vide2}) could be
considered as alternative
definitions of the zero-point energy difference, { i.e.}, 
of the vacuum energy for the theory \cite{schwinger}. 
Both definitions are 
equivalent, as we shall see, {\it only if} the effective theory is 
symmetric under CP transformation. 

Using the latter expression, the result 
is found to be even in 
the parameter $b$ \cite{kiers,border}.
Using the former one, we had 
found it to be odd \cite{border} and 
the authors of \cite{kiers,mimura} stressed that this last 
result was incomplete.  Their criticism triggered the present
study which lead us also, although with different arguments, to the
conclusion that the result should be even in $b$.

To summarize we will mainly discuss two
 issues: the proper description of the vacuum energy and the UV dependence of
 the result in order to 
clarify the differences between the approaches in 
\cite{kachelriess,kiers,border,mimura},
which, needless to repeat, all agree on the main issue: massless neutrinos do not
imply any catastrophe.

The energy density  $w(\vec x)$, the integral of which is $W$ in  Eq. (\ref{W}),
 is given in terms of Hamiltonian densities
 by 
 \bea
 w(\vec x)\equiv\langle\hat{0}|{\cal H}(\vec x)|{\hat{0}}\rangle - 
 \langle{0}|{\cal H}_0(\vec x)|{0}\rangle\ .
\label{energy} \eea
In order to compute  it 
following  Schwinger \cite{schwinger}, we write it as: 
 \bea
 w(\vec x)= -i{\partial\over\partial x^0 }
 {\mathrm{tr}}\left\{\gamma_0
 \left [ S_F( x,  y)-S_F^{(0)}(x, y)\right ]\right \}_{ y\to x}\ ;
 \label{Wanalyt}
\eea
the index $(0)$ refers to the free vacuum and $S_F$ is the usual 
 propagator, 
\bea
 S_F(x, y)\gamma_0&=&\theta(x^0-y^0)\sum\!\!\!\!\!\!\!\int _{\kappa,E>0}
 \psi_{\kappa,E}(\vec x)\psi^{\dag}_{\kappa,E}(\vec y)\,e^{-iE(x^0-y^0)}\cr
&-&\theta(y^0-x^0)\sum\!\!\!\!\!\!\!\int _{\kappa,E<0}
 \psi_{\kappa,E}(\vec x)\psi^{\dag}_{\kappa,E}(\vec y)
 \,e^{-iE(x^0-y^0)}\ , \nn \\ \label{prop}
\eea
where $H \psi_{\kappa, E} = E \psi_{\kappa,E}$, $\kappa$ 
summarizing the other quantum numbers, for instance
the momenta. 
 ~From Eq. (\ref{prop}),
 $w(\vec x)$ can be rewritten using the notation 
  $\sum \!\!\!\!\!\int\ _{\kappa,E}\
   \equiv\ \sum_{\kappa}\int dE \ $  
  in the following two ways.

  i) The first one consists in taking 
the limit  $y^0 \to x^0$ with $y^0 > x^0$: 
\bea
w(\vec x)=\sum\!\!\!\!\!\!\!\int _{\kappa,E<0} E \ \psi^{\dag}_{\kappa,E}
(\vec x)
\psi_{\kappa,E}(\vec x)  \nonumber \\
-\sum\!\!\!\!\!\!\!\int _{\kappa,E<0}E\  
\psi^{\dag(0)}_{\kappa,E}(\vec x) \psi^{(0)}
_{\kappa,E}(\vec x) \ .\label{somme}
\eea
Obviously this choice is equivalent to (\ref{vide1}).

ii) The second one corresponds to taking the symmetric average of the
 limits $y^0 \to x^0$ with 
$y^0 > x^0$
and $y^0 \to x^0$ with $y^0 < x^0$ in Eq. (\ref{prop}), as done in 
\cite{kiers} and also in
\cite{border}, which gives
\bea
 w_{{\mathrm{sym}}}(\vec x)=\frac 1 2 \sum\!\!\!\!\!\!\!\int _{\kappa,E<0} E 
 \psi^{\dag}_{\kappa,E}(\vec x)
\psi_{\kappa,E}(\vec x) \nonumber \\
-\frac 1 2\sum\!\!\!\!\!\!\!\int _{\kappa,E<0}E
\psi^{\dag(0)}_{\kappa,E}(\vec x) \psi^{(0)}
_{\kappa,E}(\vec x) \nonumber \\
-\frac 1 2 \sum\!\!\!\!\!\!\!\int _{\kappa,E>0} E
\psi^{\dag}_{\kappa,E}(\vec x)
\psi_{\kappa,E}(\vec x) \nonumber \\
+\frac 1 2\sum\!\!\!\!\!\!\!\int _{\kappa,E>0}E
\psi^{\dag(0)}_{\kappa,E}(\vec x) \psi^{(0)}
_{\kappa,E}(\vec x).\label{sommesym}
\eea
corresponding to (\ref{vide2}).
 The symmetrization over the limit on the time components is equivalent to the
 symmetrization over positive and negative energies as a consequence of time boundary
 conditions of the usual Feynman propagator\footnote{It is worth pointing out that the use of
 momentum-space Feynman propagators to rewrite the energy density, $w(\vec x)$, implies
 an implicit symmetrization over positive and negative energies because of the Fourier
 integration over the time and of the time boundary conditions. Thus, in Feynman's picture
 the symmetrization is assumed.}. Moreover, this symmetrization leads to an even
 behaviour in $b$: if one takes into account the transformation of the
 Lagrangian under CP, $CP
{\cal {L}}(b) (CP)^{-1} = {\cal {L}}(-b)$, Eq. (\ref{sommesym}) can be rewritten as

 \bea
 w_{{\mathrm{sym}}}(b, \vec x)={1\over 2} \sum\!\!\!\!\!\!\!
 \int _{\kappa,E<0}E \left\{ 
 \psi^{\dag}_{\kappa,E}
\psi_{\kappa,E}(b,\vec x)+ \nonumber \right. \\
\left. \psi^{\dag}_{\kappa,E}
\psi_{\kappa,E}(-b,\vec x) -2\psi^{\dag}_{\kappa,E}
\psi_{\kappa,E}(0,\vec x) \right\} \ \ ; \label{even}\eea 

\noindent where the dependence on $b$ of the eigenfunctions is explicitly
written.

 In Refs. \cite{kiers,border} this
result has been confirmed, while by using the non-symmetric 
expression (\ref{somme}) the
energy was linear in $b$ \cite{border}. In order to understand
 this difference, 
we can subtract Eq. (\ref{sommesym}) from (\ref{somme}):
\bea w(\vec x)-w_{{\mathrm{sym}}}(\vec x)= \left\{ \frac i 2 \frac \partial 
{\partial x_0} 
\sumint_{\kappa,E} \left( \psi^{\dag}_{\kappa,E}(\vec y)
\psi_{\kappa,E}(\vec x)  \right. \right. \nonumber \\
\left. \left. -\psi^{\dag \ (0)}_{\kappa,E}(\vec y)
\psi^{(0)}_{\kappa,E}(\vec x) \right) \ e^{-iE(x_0-y_0)} \right\}_{
y \to x} \nn \\
= \ \frac i 2  \frac \partial {\partial x_0}
 \delta_{\alpha,\beta}\left\{ \langle \hat{0} | \left[
\Psi^{\dag}_{\alpha}(y),\Psi_{\beta}(x) \right]_{+} | \hat{0} \rangle 
\right. \nonumber \\
\left. - \langle 0 | \left[ \Psi^{\dag \ (0)}_{\alpha}(y),
\Psi^{(0)}_{\beta}(x) \right]_{+} | 0 \rangle 
\right\}_{y \to x} \ \ , \label{canonic}
\eea

\noindent where $[\Psi^{\dag},\Psi]_{+}$ stands for the canonical
anticommutation relation of the fermionic field:
\bea
\Psi(x)= \sumint_{\kappa, E>0} \ \psi_{\kappa,E}(\vec x) \ e^{-iEx_0}
\ b_{\kappa,E} \nonumber \\
+ \ \sumint_{\kappa', E<0} \ \psi_{\kappa',E}(\vec x) 
e^{-iEx_0} d^{\dag}_{\kappa',E} 
\eea
\noindent the ``coefficients'' $b_{\kappa,E}$ and $d^{\dag}_{\kappa,E}$ being the usual
fermionic annihilation and antifermionic creation operators; $\kappa'$ 
is obtained from $\kappa$ by performing
the  canonical transformation generating the
positive-energy antiparticle states from the negative-energy ones. 

 From the CP transformation:
$\Psi(x_0,\vec{x}) \ \to  \ i\gamma_2 \gamma_0 \Psi^\ast(x_0,-\vec{x}) 
\ $\cite{itzykson}, it is
easy to see that the last r.h.s of (\ref{canonic}) changes sign under CP.
  Therefore 
(\ref{canonic}) vanishes {\it if the vacua} ( $|\hat{0}\rangle,|0\rangle$) {\it
are CP-invariant}. An equivalent way of seeing this is to remark that
in a CP-symmetric vacuum, there is  an $(E,\kappa \to
-E,\kappa')$-symmetry, inducing a cancellation in the sums of the first r.h.s in
 (\ref{canonic}). It results that in a CP-invariant vacuum, (\ref{vide1}) and
 (\ref{vide2}) or equivalently  (\ref{somme}) and (\ref{sommesym}) are equal. 

In a {\it non-CP-symmetric vacuum} $|\hat{0}\rangle$, namely a  neutronic 
vacuum leading to the non-CP-invariant effective Lagrangian for the neutrinos
(\ref{lageff}), which has been  used by the authors of Refs.
\cite{rescue,kachelriess,note,kiers,border,mimura}, Eq. (\ref{canonic})  {\it does not vanish.} 
Indeed, the axial charge (to which the parameter $b$ is proportional) changes
sign under CP transformation\cite{itzykson} and, in fact, 
our effective Lagrangian is
only invariant under the product of CP (which changes neutrinos into
antineutrinos) and the operation that  changes $b$ 
into $-b$ (which changes the neutron star into an antineutron star, 
as mentioned by Kiers et Tytgat in \cite{kiers}). 

\begin{figure}[hbt]   % produce figure here
\begin{center}
\mbox{\epsfig{file=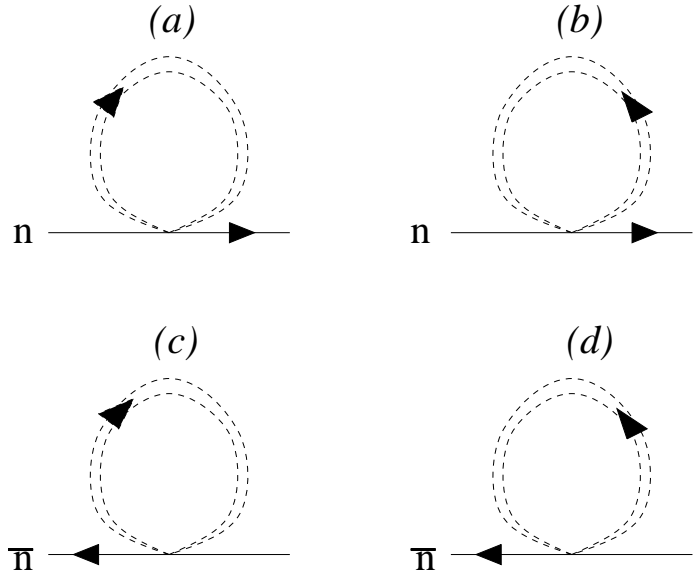,height=8cm}}
\caption[]{\it{ The diagram (a) is changed to (d) and (b) to (c) by
reversing the time arrow. In the Feynman's picture, reversing the time
arrow implies a change of fermionic lines into  antifermionic lines. Thus,
CP invariance guarantees that the
total result will remain the same for both  neutron and antineutron
stars, (a)+(b)=(d)+(c). The  dashed double 
line represents  the neutrino propagator in the medium of the star.}} 
 \protect\label{ps}
\end{center}
\end{figure}

Therefore CP invariance\footnote{In this note we neglect CP violation in the
Standard Model. Still, if the neutron stars are considered at equilibrium, it
suffices to invoke CPT to impose an equal mass to neutron and antineutron
stars.}  
imposes the invariance of the ``vacuum'' energy    under the exchange 
$b\ \to \ -b $: the energy due to massless neutrino multibody exchange inside a 
neutron star is the 
same as the one inside an antineutron star  
(see  Fig. \ref{ps}).

In principle, the problem of  matching  the effective theory with the
 underlying one is related to the appropriate choice of the ultraviolet 
 regularization. 
 A detailed and faithful description of the transition from the 
 underlying theory to the effective theory would lead to a UV-regularization, 
 bringing automatically into the effective theory the wanted properties such
 as the CP symmetry. 
 However, this is a very complex task. The limited goal faced
in Refs. \cite{rescue,kachelriess,note,kiers,border,mimura} justifies a naive 
``{\it description}'' of 
UV physics; for instance, the use of a simple
cut-off in Ref. \cite{border}. We will hence invoke 
the general symmetry
properties of the underlying theory (CP) to constrain the description of the
vacuum energy in the effective theory (\ref{vide2}), exactly as the chiral symmetry of QCD constrains
the effective chiral Lagrangian.

Eq. (\ref{even}), and hence (\ref{sommesym}), is clearly invariant  
under $ b\ \to \
-b$ transformation. The latter 
equation is no longer equivalent to (\ref{somme}) since the  difference given
 by Eq. (\ref{canonic})
is non-zero, the Dirac equation obtained from (\ref{lageff}) is not
symmetric in the exchange $E,\kappa \ \to \ -E,\kappa'$.
Thus, the CP symmetry of the underlying theory, i.e. Q.C.D., imposes the choice
 of 
(\ref{sommesym}) in order to compute the interaction energy density. 

 For practical calculations we need to remember that our effective 
 theory  is valid only 
 under the assumption
 that the neutrons are static and remain so until the energy scale $C\sim 100
 $ MeV. Moreover, there are other energy scales that limit our theory, as 
 the one related to
 the confinement of QCD ($\sim 1 $ GeV), the one related to the 
 mean free path of neutrinos 
 in the star... The physics related to
 these UV cut-offs has been  discussed in  Ref. \cite{border}.
 The use of a cut-off in the energy integration can be alternatively 
 interpreted as
 the assumption
  that a repulsive core prevents the neutrons 
  from ``piling-up" in space, as noticed by
  Fischbach \cite{fischbach}. 
  In our effective theory  of Eq. (\ref{lageff}), 
  $b$ can be naively interpreted as the coherent and homogeneous  amplitude 
  for static neutrons to interact with 
  neutrinos. The energy cut-off expresses 
  the scale at which neutron recoil, repulsive cores, quark and gluon
  substructures, etc., induce the neutrino interaction with non-static neutrons
   to become incoherent and inhomogeneous.

  In \cite{border}, using one of these physical cut-offs, $C$, we have found
   for the energies, using respectively Eqs. (\ref{vide1}) and (\ref{vide2}):
 \bea  
 W\sim
-{b C^3}{R^3} .\label{faux} \eea
\bea
 W_{{\mathrm{sym}}}
  \sim - {b^2 C^2}{R^3}\ .\label{estimate-sym}
 \eea
Eq. (\ref{estimate-sym}), which uses (\ref{vide2}), {\it is compatible with
the CP symmetry of QCD}.  
Beside the cut-off explicit regularization, it is probable  that the result should also depend
on the way of performing the
summation over the energy. Indeed, this is the first step in the regularization process
\cite{note}.

In computing $W_{{\mathrm{sym}}}$, Kiers and Tytgat \cite{kiers} counted the energy levels of the 
Hamiltonian by putting the
system in a large box,  which was a  method different from the one in \cite{border}, and their result was
\bea  
 W_{{\mathrm{sym}}}^{\mathrm KT}
 \sim - {b^4 }{R^3}.\label{kt}
 \eea

Let us present a few comments to understand the difference between
(\ref{estimate-sym}) and (\ref{kt}). Both are even in $b$ and respect CP
symmetry.
Next, the sums leading to (\ref{estimate-sym})   and to (\ref{kt})  
correspond to different
orderings  of the same energy levels. Reordering a sum would not change
the result if we were not speaking of divergent series
\footnote{The classical example is that of the two possible
orderings of the non-convergent series $(-1)^n$:
 $\sum_n (-1)^n=1+(-1+1)+(-1+1)+\cdots$ and  
 $\sum_n (-1)^n=(1-1)+ (1-1)+\cdots$, which are both valid to describe 
 the series.}, which have been
regularized in  ways that turn out to be different. This can be
illustrated simply   by applying
 both summation methods (\cite{border} and \cite{kiers}) to the
  ($1 \ + \ 1$)-dimensional toy 
 model we presented in Ref. \cite{note}: we  found
 that $W=0$
 because there was a one-to-one correspondence between the energy levels inside
  and outside the star; by putting the $(1\ +\ 1)$ star in a box, as done in
   Ref. \cite{kiers},
  we verified that the result was not zero, although it was extremely small.
 So the way of summing over the energy levels gives a difference in the final
  result. 
  
  Furthermore, the main difference between the two  symmetrized results, Eqs.
  (\ref{estimate-sym}) and (\ref{kt}), is the power in the parameter $b$. The
  origin of that discrepancy comes from  neglecting the two-body
  contribution to the energy density performed by the authors of \cite{kiers}. 
  They
  argue that the two-body interacting energy behaves as $R^2$, 
  its contribution
  to the energy density scales as $1/R$ and this becomes  negligible in the
   large $R$-physical regime. This seems to contradict 
  previous studies that supported  the existence of a well-defined neutrino 
  exchange
  two-body potential\cite{Fein}
  \bea
  {G_F^2\over 4\pi^3} {1\over|\vec r_1-\vec r_2|^5}\ . \label{potential}
  \eea
  The integration of the latter potential over space is UV-divergent. Some
  regularization procedures lead to a vanishing result,  for
  example, Pauli-Villars in \cite{rescue,note}, and
  dimensional regularization in \cite{kiers}. 
 This vanishing of the integrated two-neutrino {\it positive}
   potential (\ref{potential}) is physically surprising, and is
equivalent to assuming a negative distribution
   located at $r_1=r_2$. Another regularization, such as a simple UV
cut-off: $|\vec r_1-\vec r_2| > r_c$
   certainly leads to a positive non-vanishing result $\propto 1/r_c^2$. 
   
  If one still accepts this vanishing \cite{kiers}, then no $b^2$-terms
  remain, and only diagrams with at least four neutrons 
  contribute; those contributions start at $b^4$.  However, in
   Ref. \cite{kiers}, the authors stated that for a more realistic star, the
   results should be proportional to $b^2$. 
 
  Finally, in Ref. \cite{border} we use a simple cut-off as
  regularization procedure, obviously different from both 
  the {\it Pauli-Villars} and the 
   {\it dimensional} scheme used, respectively, by us in Refs. \cite{rescue,note} and by
  Kiers and Tytgat in Ref. \cite{kiers}; we then  perform 
  a {\it ``crude''} approximation, 
  retaining only the sharp effects of the neutrino potential. Our result is
  proportional to $b^2$ (Eq. (\ref{estimate-sym})). We might wonder why this ``crude''
  result does not show larger powers of $b$. We may conjecture that this is due 
  to the fact the two-body potential, being the shortest range one,  is
  the one that  corresponds to the ``sharp'' effects considered in the ``crude''
  approximation, while the many-body potentials, being long-range, contribute to
  the corrections to the ``crude approximation'' studied in \cite{border}.
 
  As a conclusion we believe the result (\ref{estimate-sym}) to be quite
  reasonable. However, our
  main conclusion is that the energy should  be even 
 in $b$ because of
CP invariance and that the symmetrization given by Eq. (\ref{sommesym}) is
required.   
 That the resulting energy is even in $b$ is also manifest in Schwinger's
 expansion \cite{fischbach}, as it results from the symmetrization (Furry's
 theorem) implicitly assumed in a Feynman diagram treatment of the problem.

\vskip 1.5cm
\bigskip

We are specially indebted to  M. B. Gavela for helpful discussions.
 We wish to thank K. Kiers and M. Tytgat
 for very important comments about that question.
 This work has been partially supported by Spanish CICYT, project 
PB 95-0533-A, and  by the Fundaci\'on Ram\'on Areces.

\end{document}